\documentclass[12pt]{article} 
\usepackage{graphicx,subfigure,latexsym,hyperref,color}

\newcommand{\be}{\begin{equation}}
\newcommand{\ee}{\end{equation}}
\newcommand{\bear}{\begin{eqnarray}}
\newcommand{\eear}{\end{eqnarray}}
\newcommand{\ba}{\begin{array}}
\newcommand{\ea}{\end{array}}

\newcommand{\mc}{\mathcal}

\begin{document}


\vfill
\title{\bf \Large{Triangle anomaly in Weyl semimetals}}
\date{\today}
\author{G\"ok\c ce Ba\c sar$^{1}$\footnote{e-mail: {\tt basar@tonic.physics.sunysb.edu}}, Dmitri~E.~Kharzeev$^{1,2}$\footnote{e-mail:
{\tt dmitri.kharzeev@stonybrook.edu}},and Ho-Ung Yee$^{3,4}$\footnote{e-mail:{\tt hyee@uic.edu}} 
\\ \ \\
{\it $^{1}$Department of Physics and Astronomy, Stony Brook University,} \\
{\it Stony Brook, New York 11794-3800 }\\[0.1in]
{\it $^{2}$Department of Physics, Brookhaven National Laboratory,} \\
{\it Upton, New York 11973-5000 }\\[0.1in]
{\it $^{3}$Department of Physics, University of Illinois, Chicago, Illinois 60607}\\[0.1in]
{\it $^{4}$RIKEN-BNL Research Center, Brookhaven National Laboratory,}\\
{\it Upton, New York
11973-5000}}

\maketitle
\vskip-13cm\hskip13cm RBRC--1024
\vskip13cm
\vfill

\begin{abstract}

Weyl semimetals possess massless chiral quasi-particles, and are thus affected by the triangle anomalies.
We discuss the features of the chiral magnetic and chiral vortical effects specific to Weyl semimetals, and 
then propose three novel phenomena caused by the triangle anomalies in this material: 
1) anomaly cooling; 2)  charge transport by soliton waves as described by the Burgers' equation, and 3) the shift of the BKT phase transition
of superfluid vortices coupled to Weyl fermions. In addition, we establish the conditions under which the chiral magnetic current exists in real materials.

\end{abstract}

\vfill


%


\setcounter{footnote}{0}

\baselineskip 18pt \pagebreak
\renewcommand{\thepage}{\arabic{page}}
\pagebreak

\section{Introduction}

The triangle anomaly is a quantum effect responsible for the violation of chiral symmetry of charged massless fermions 
in the presence of a P- and CP-odd configuration of the background gauge field \cite{Adler:1969gk,Bell:1969ts}. A massless Dirac fermion, such as a (nearly) massless quark in QCD, 
possesses a left- or right-handed chirality. These states are described by the left- and right-handed Weyl spinors giving rise to the  
$U(1)_L$ and $U(1)_R$ chiral symmetries. The fermions with different chiralities contribute to the triangle anomaly with 
opposite signs -- as a result, the anomaly is absent for the vector current $J_V\equiv J_L+J_R$, and the electric charge is conserved. 
On the other hand, for the axial current $J_A\equiv -J_L+J_R$ it leads to 
\be
\partial_\mu J^\mu_A = {e^2\over 2\pi^2} \vec E\cdot \vec B\quad,
\ee
where $e$ is the charge of the fermion; the sum over different fermion species is implicit here; note that this is the covariant form of anomaly.
It will be useful for our purposes to view the above equation in terms of two separate anomaly equations for the left- and right-handed chiral
currents:
\be
\partial_\mu J^\mu_{L,R} =\mp{e^2 \over 4\pi^2} \vec E\cdot\vec B\quad.\label{LRanomaly}
\ee

Triangle anomaly plays an important role in the chiral dynamics of low energy QCD, explaining in particular 
the $\pi^0\to \gamma\gamma$ decay. Recently, it has become clear that the anomaly also affects the transport and hydrodynamical macroscopic behavior; 
much of this work was motivated by the applications to quark-gluon plasma in magnetic field \cite{Kharzeev:2007jp,Fukushima:2008xe} 
produced in heavy ion collisions. The novel transport phenomena induced by the anomaly include the Chiral Magnetic Effect (CME) \cite{Kharzeev:2007jp,Fukushima:2008xe,Kharzeev:2004ey, Kharzeev:2007tn}, the chiral separation effect \cite{son:2004tq,Metlitski:2005pr}, and 
the Chiral Vortical Effect (CVE)  \cite{Kharzeev:2007tn,Vilenkin:1979ui, Erdmenger:2008rm,Banerjee:2008th,son:2009tf}.

The CME refers to the electric current along an external magnetic field induced by the chirality imbalance. Because the electric current is a vector, and magnetic field -- a pseudo-vector, the CME is a parity--odd phenomenon. The CVE is an analogous effect induced by the presence of vorticity and chirality imbalance, at finite chemical potential. In the context of condensed matter physics, chirality emerges in the vicinity of the band touching points where the quasi-particle dispersion relation is linear and the quasi-particle is described by the Weyl spinor. Closely related phenomena have been discussed in the physics of neutrinos  \cite{Vilenkin:1979ui}, conductors with mirror isomer symmetry \cite{elia,leonid}, primordial electroweak plasma \cite{Giovannini:1997gp} and quantum wires \cite{acf}. Note that the axial anomaly and the topology of background gauge field are crucial for the existence of the chiral magnetic current; without the anomaly, this current has to vanish in thermal equilibrium. The possible existence of CME in Weyl semimetals has been discussed previously in \cite{cme,zyuzin_burkov,Goswami,Grushin:2012mt,Franz}). 

For systems that contain charged chiral fermions, the chiral magnetic (separation) effects dictate the existence of vector (axial) charge currents along the direction of an external magnetic field in the presence of the axial (vector) chemical potential. In the case of the chiral vortical effect, the role of magnetic field is played by the vorticity of the fluid. Due to the topological nature of triangle anomaly,  these new transport 
phenomena have been shown to be robust and not modified by interactions even in the strong coupling limit \cite{Yee:2009vw,Rebhan:2009vc,Gynther:2010ed}, with a possible exception \cite{Golkar:2012kb,Hou:2012xg} of the temperature-dependent $\sim T^2$ term \cite{Landsteiner:2011cp,Neiman:2010zi} in the chiral vortical conductivity. 

The persistence of the anomalous charge transport at strong coupling suggests the possibility of hydrodynamical formulation, and such formulation was given in Ref \cite{son:2009tf}, see also \cite{Lublinsky:2009wr,Sadofyev:2010is,Lin:2011aa,Nair:2011mk}. The absence of contributions to the local entropy production rate from the (T-even) anomalous terms has been used to constrain the hydrodynamical formulation \cite{Kharzeev:2011ds}. In heavy ion collisions, the chiral magnetic and chiral vortical effects can potentially be separated by measuring the electric charge and baryon number asymmetries \cite{Kharzeev:2010gr,Rogachevsky:2010ys}. The experimental evidence 
for the chiral magnetic effect in heavy ion collisions has been presented by RHIC \cite{Abelev:2009ac,Abelev:2009ad,Wang:2012qs} and LHC \cite{Selyuzhenkov:2012py,Abelev:2012pa,Hori:2012hi} experiments.


Recently, it has been realized that the triangle anomalies and the chiral magnetic effect can be realized also in a condensed matter system -- 
a (3+1)-dimensional Weyl semimetal \cite{Wan:2011,wsm,cme}. The existence of ``substances intermediate between metals and dielectrics" with the point touchings of the valence and conduction bands in the Brillouin zone was anticipated long time ago \cite{Abrikosov}. 
In the vicinity of the point touching, the dispersion relation of the quasiparticles is approximately linear, as described by the Hamiltonian 
$
H = \pm v_F \vec{\sigma} \cdot \vec{k},
$ 
where $v_F$ is the Fermi velocity of the quasi-particle, $\vec{k}$ is the momentum in the first Brillouin zone, and $\vec{\sigma}$ are the Pauli matrices. 
This Hamiltonian describes massless particles with positive or negative (depending on the sign) chiralities, e.g. neutrinos, and the corresponding wave equation is known as the Weyl equation -- hence the name {\it Weyl semimetal} \cite{Wan:2011}. Weyl semimetals are closely related to 2D graphene \cite{Geim:2007}, and to the topological insulators \cite{TI,Volovik} -- 3D materials with a gapped bulk with nonzero Berry fluxes  and a surface supporting gapless edge excitations.  Specific realizations of Weyl semimetals have been proposed, including doped silver chalcogenides $Ag_{2+\delta}Se$ and $Ag_{2+\delta}Te$ \cite{Abrikosov1}, pyrochlore irridates $A_2 Ir_2 O_7$ \cite{Wan:2011}, and 
a multilayer heterostructure composed of identical thin films of a magnetically doped 3D topological insulator, separated by ordinary-insulator spacer layers \cite{wsm}. 

The triangle anomaly affects Weyl semimetals \cite{Nielsen:1983rb,Volovik,Son:2012wh} because the fermionic quasi-particles around the Weyl point in momentum space behave like relativistic chiral fermions
with a velocity that plays the role of an effective speed of light \cite{Wan:2011,wsm,cme,weyl3}. However, as observed in Ref.\cite{Volovik,Son:2012wh}, 
the crux of triangle anomaly in Weyl semimetals is the presence of a hedgehog, or a magnetic monopole, in momentum space, that leads to the emergence of Berry's phase \cite{berry} (see also Ref.\cite{Zahed:2012yu}).  The Berry's phase, and the anomaly, thus can affect the systems that are not truly relativistic.  

If the total flux of Berry's phase is an integer $k$, it induces a non-conservation of the charged  fermion current through the triangle 
anomaly:
\bear
\partial_\mu J^\mu = {k e^2\over 4\pi^2}\vec E\cdot\vec B\quad,\label{cmanomaly}
\eear
where $e$ is the charge of the quasi-particles. The analogy to (\ref{LRanomaly}) is clear; each Weyl point with a monopole charge $k$ is similar to a relativistic chiral species with chirality dependent on the sign of $k$.
The total number of electrons in the system should be conserved, and the sum of monopole charges $k$ over all Weyl points
must thus be zero.

As the newly discovered transport phenomena mentioned above rely only on the triangle anomaly relation (\ref{cmanomaly}),
they should be present also in Weyl semimetals. This is important as it would allow to test the transport phenomena originating from the triangle anomaly
experimentally in a controlled environment. 
Our purpose in this paper is to provide a few examples that may have potential experimental or even practical importance; 
see Refs.\cite{Son:2012bg,Gorsky:2012gi} for previous suggestions and Ref.\cite{Kharzeev:2012dc} for a discussion of ``chiral electronics" enabled by Weyl semimetals.

The new transport phenomena on which we base the subsequent discussion can be summarized as follows. 
Each Weyl point with $k$ total flux of Berry's phase contributes to the electromagnetic current through the relation \cite{Fukushima:2008xe,Vilenkin:1979ui,son:2009tf,Landsteiner:2011cp,Neiman:2010zi}
\be
\vec J={k e^2\over 4\pi^2 } \mu \vec B+{k e \over 4\pi^2}\left(\mu^2+{\pi^2 \over 3}T^2\right) \vec \omega\quad,\label{current}
\ee
where $e$ is the electric charge
of the fermionic quasi-particles of the Weyl point, $\mu$ is the chemical potential of the Fermi surface measured from the Weyl point, and $T$ is the temperature.
The magnetic field $\vec B$ and the vorticity $\vec \omega={1\over 2}\vec \nabla\times\vec v$ of the velocity field $\vec v$
should be computed in the local rest frame.
In addition, it has been argued that the transport phenomena originating from triangle anomaly do not lead to entropy production; they
are non-dissipative \cite{Kharzeev:2011ds}. In Refs.\cite{Kharzeev:2011ds,Loganayagam:2011mu,Loganayagam:2012pz} the non-dissipative nature of anomalous currents and the resulting time-reversal invariance of anomalous conductivities have been exploited to extend the 
first order anomalous hydrodynamics \cite{son:2009tf} 
 to second order in derivatives and to higher dimensional cases.
The entropy current originating from the anomaly is given by \cite{son:2009tf}
\be
\vec S=ke\left({1\over 8\pi^2}{\mu^2\over T}+{T\over 24}\right)\vec B+k\left({1\over 12\pi^2}{\mu^3\over T}+{\mu T\over 12}\right)\vec\omega \quad,\label{entropy}
\ee
where the $\vec B$ and $\vec \omega$ should be computed in the local rest frame.

\section{Chiral magnetic and chiral vortical effects in Weyl semimetals}
\subsection{CME and the conditions for its existence}
Prior to discussing possible experimental consequences of the anomaly-induced transport, let us make a few cautionary remarks on the application of (\ref{current}) to real Weyl semimetals where the electron energy spectrum differs from that  of free relativistic chiral fermions.

\begin{figure}[t]
	\centering
	\includegraphics[scale=0.29]{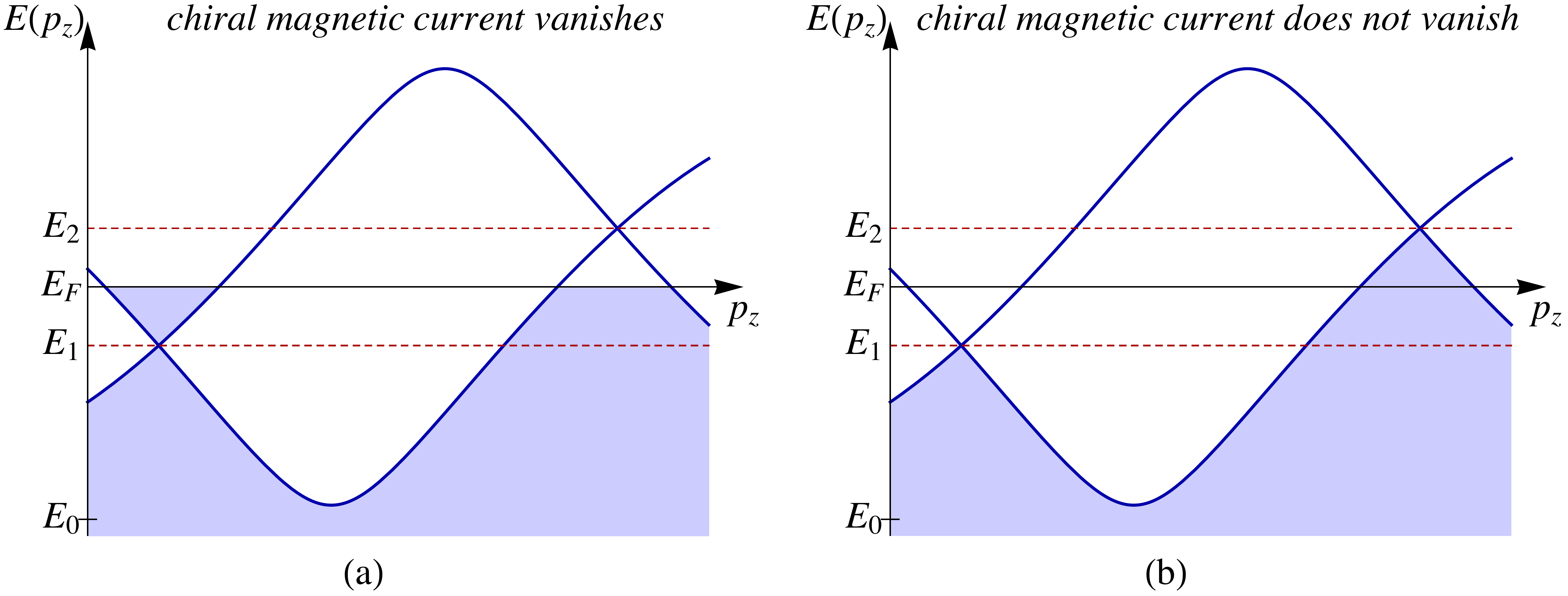}
		\caption{A schematic illustration of the energy dispersion relation of a Weyl semimetal considered in Ref.\cite{cme} . The existence of the chiral magnetic current depends on how the levels are filled as denoted by the shaded regions. Fig. 1a corresponds to the absence of the chiral chemical potential, and thus the absence of the chiral magnetic current. In contrast, Fig. 1b describes the situation with a non-zero chiral chemical potential, in which the chiral magnetic current exists. } 
\label{fig0}
\end{figure}

At first glance it appears that
 one can have a net CME even in global equilibrium if the energies of Weyl points are shifted in an asymmetric way by introducing an inversion-symmetry breaking term. 
Since the energy of each Weyl point, say $E_i$, is now shifted away from the Fermi energy $\varepsilon_F$, it seems naively that each Weyl point has an effective chiral chemical potential $\mu_i=\varepsilon_F-E_i$ measured from the origin (at zero temperature) even in global equilibrium when the bands are filled up to the Fermi energy $\varepsilon_F$, see 
Figure \ref{fig0}. 
If one naively applies (\ref{current}) using these chiral chemical potentials $\mu_i$, a net chiral magnetic effect would result.

However, the existence of the chiral magnetic current in global equilibrium (or ``vacuum'') at zero temperature would raise a conceptual issue.
Suppose one applies a parallel electric field in addition to the magnetic field; because of the chiral magnetic current, there would be net energy input (or output),
\be
{dP\over dt}=\vec J\cdot \vec E \sim \vec E\cdot\vec B\quad,
\ee
where the sign can be made negative by choosing $\vec E$ appropriately -- that is, one could extract energy out of the system.
However, it should be impossible to extract energy from a state in global equilibrium (``vacuum'') at zero temperature since the state is already
a minimal energy state by definition; there is simply no energy available. This argument dictates 
the absence of the chiral magnetic current in the equilibrium, by which we mean the configuration where both left- and right-handed sectors are filled up to the {\it same Fermi energy} as shown in Figure \ref{fig0} (a). The same argument applies not only to Weyl semimetals, but also to neutrinos in magnetic field 
\cite{Vilenkin:1979ui}, and to conductors with mirror-isomeric structure \cite{elia}-- in all of these cases, the chiral magnetic current has been eventually found to vanish in equilibrium \cite{Vilenkin:1979ui,leonid}. 

On the contrary, if each chiral sector is filled up to a different energy level, as shown in Figure \ref{fig0} (b), then the chiral magnetic current does not vanish. The magnitude of the chiral magnetic current is proportional to the difference between the energies up to which each sector is filled, i.e. the chiral chemical potential. The existence of the CME current further requires that the chiral chemical potential can evolve as a function of time. The chiral magnetic effect discussed in \cite{Fukushima:2008xe} is based on the assumption that the chiral chemical potential is nonzero and is not fixed, i.e. the system is not in the minimal energy state. The chiral magnetic current is thus powered by the energy stored in the difference of the Fermi energies of the left- and right-handed chiral fermions \cite{Kharzeev:2009fn}. 
If this difference is fixed,
the chiral magnetic current cannot exist. 

In this section, we will establish the conditions necessary to realize the chiral magnetic effect in a Weyl semimetal.
Our results indicate that (\ref{current}) is {\it almost} correct in realistic Weyl semimetal systems, but with some subtle
modifications that we will discuss in detail. With these modifications, we find that the system in Figure \ref{fig0} (a)
has no net chiral magnetic effect (CME) in global equilibrium. However, the CME can nevertheless be realized if each sector is filled up to a different energy as shown in Figure \ref{fig0} (b). An example of the latter configuration can be realized if the chiral chemical potential is generated dynamically, e.g. by ``chirally charging" the Weyl semimetal in parallel electric and magnetic fields. Other realizations may also be possible. 

For our purposes it will be sufficient to use 
a kinetic approach developed in Refs.\cite{Son:2012wh,sundaram,haldane,Gao:2012ix,Stephanov:2012ki,Stone:2013sga}. 
In kinetic approach, one assumes that the quasi-particle phase space distributions are described classically by the Boltzmann equation, and the 
collisions between the quasi-particles (Weyl fermions in our case) are rare. The effect of the chiral anomaly, which is a quantum phenomenon, is captured by an additional term in the action of kinetic theory: the Berry phase. For free Weyl fermions, the Berry phase has the form of a magnetic monopole \cite{berry, book} in momentum space which can be expressed as
\be
S_{Berry}=-\int dt\, {\bf a}({\bf p})\cdot\dot{\bf{p}}
\ee
where $\bf{a}(\bf p)$ has the form of a gauge potential of a magnetic monopole. Specifically, we will rely on the formulation of Ref.\cite{Stephanov:2012ki}. 
Each $i$'th Weyl point of Berry monopole charge $k_i$ is assumed to be situated at the energy $E_i$ and the momentum $\vec k_i$, so
that the dispersion relation around that point is approximately linear:
\be\label{branch1}
E(\vec p)= E_i\pm v_i|\vec p-\vec k_i|\quad.
\ee
It is important to recognize that the states with positive energy (upward from $E_i$, the branch with the positive sign in (\ref{branch1})) feel the Berry phase of charge $k_i$ while the states with negative energy (downward from $E_i$) feel the opposite charge $-k_i$.
(It is clear from the example of $k=1$ with the Hamiltonian $H=E_i 1+v_i\vec\sigma\cdot(\vec p-\vec k_i)$).

Note that we have not introduced any notion of ``holes'' or anti-particles; it is an important point that deviates from the usual discussion of relativistic chiral fermions.
In the context of condensed matter physics, it is clearly natural as the fermions in the filled ``Dirac sea'' are physical electrons.
In the relativistic chiral fermion description, one {\it could} treat the Dirac sea fermions as electrons with negative energy,
but the real difference comes in the fact that one should then subtract the vacuum (all negative states are occupied) contribution. This is because the only observable physical effect should originate from a difference from the vacuum contribution. 
By the identity
\be
f(-|E|)=1-\bar f(|E|)\quad,
\ee
where $f(E)$ is the distribution of particles at all energy $E\in (-\infty,+\infty)$ and $\bar f(E), E>0$ is the anti-particle distribution, one can show that this 
gives the equivalent results to the anti-particle framework.
In our condensed matter situation, we can no longer {\it a priori} know the results for the ``vacuum''
(it is precisely this question that we are addressing now), so treating both positive and negative branches as electrons will allow to avoid confusion.

For the positive branch, the quasi-electrons are described by the action \cite{Son:2012wh,Stephanov:2012ki},
\be
I_+=\int dt\,\left( {\bf p}\cdot\dot{\bf{x}} +{\bf A(x)}\cdot\dot{\bf{x}} -E_i-v_i|{\bf p}-{\bf k}_i|-{\bf a}({\bf p})\cdot\dot{\bf{p}}\right),\label{actionp}
\ee
where we have the Berry monopole
\be
{\bf b}\equiv {\bf \nabla}_p\times {\bf a} = {k_i({\bf p}-{\bf k}_i)\over 2|{\bf p}-{\bf k}_i|^3}\quad.\label{bb}
\ee
Following the steps in Ref.\cite{Stephanov:2012ki}, the equations of motion become
\bear
 \sqrt{G_+}\dot{{\bf x}}&=&{v_i({\bf p}-{\bf k}_i)\over |{\bf p}-{\bf k}_i|}+{k_i v_i\over 2}{{\bf B}\over  |{\bf p}-{\bf k}_i|^2}\quad,\nonumber\\
\sqrt{G_+}\dot{{\bf p}}&=& v_i {({\bf p}-{\bf k}_i)\times {\bf B}\over|{\bf p}-{\bf k}_i|}\quad.\label{particleeq}
\eear
Here $\sqrt{G_+}=(1+{\bf b}\cdot{\bf B})$ is the phase-space measure generated by the Berry phase and reflects the effect of the anomaly,
and the anomaly-induced current is
\bear
{\bf j}_+&=&\int {d^3 p\over (2\pi)^3} f(E) \sqrt{G}\dot{\bf x} ={\bf B}{k_i\over 4\pi^2}\int f(E){v_i\over |{\bf p}-{\bf k}_i|^2}|{\bf p}-{\bf k}_i|^2 d|{\bf p}-{\bf k}_i|\nonumber\\&=&
{\bf B}{k_i\over 4\pi^2}\int f(E)v_i d|{\bf p}-{\bf k}_i|={\bf B}{k_i\over 4\pi^2}\int_{E_i}^{\infty} f(E)dE\quad,\label{parres}
\eear
where in the last equality, we use $E=E_i+v_i|{\bf p}-{\bf k}_i|$. The distribution function $f(E)$ in global equilibrium is 
\be
f(E)={1\over 1+e^{\beta (E-\varepsilon_F)}}=-{1\over\beta}{\partial\over\partial E}{\rm log}\left(1+e^{-\beta (E-\varepsilon_F)}\right)\quad,\label{disteq}
\ee
but the above formula is applicable to more general out-of-equilibrium situations as well \cite{Stephanov:2012ki}.
In the $E$ integration, the upper limit in our situation is not infinite and has a  physical cutoff.
However, this cutoff is not of significance at low enough temperature as these high energy states are rarely occupied, $f(E)\ll 1$. The above result is a reproduction of previous computations in Refs.\cite{Son:2012wh,Stephanov:2012ki}. 
We emphasize that although we assume a specific linear dispersion relation to derive the result, the final expression in (\ref{parres}) can be shown to be universal and is not sensitive 
to the detailed shape of the dispersion relation.

The small region around the origin where $\sqrt{G}=(1+{\bf b}\cdot{\bf B})\le 0$ is the quantum region \cite{Stephanov:2012ki} where the kinetic approach breaks down. Its size scales linearly in ${\bf B}$ so it gives rise to a small
correction (if any) to (\ref{parres}) of higher orders in $B$ in the small $B$ limit that we assume.

The interesting part is the negative energy branch (note again that we do not have anti-particles or holes).
The action describing these quasiparticles is
\be
I_-=\int dt\,\left( {\bf p}\cdot\dot{\bf{x}} +{\bf A(x)}\cdot\dot{\bf{x}} -E_i+v_i|{\bf p}-{\bf k}_i|+{\bf a}({\bf p})\cdot\dot{\bf{p}}\right),
\ee
where we have changed signs in two places compared to (\ref{actionp}); the momentum dependent part of the energy
according to the negative branch and the Berry phase are reversed as discussed before, with the same definition of ${\bf b}$ as in (\ref{bb}).
Similar steps lead to the equations of motion,
\bear
 \sqrt{G_-}\dot{{\bf x}}&=&-{v_i({\bf p}-{\bf k}_i)\over |{\bf p}-{\bf k}_i|}+{k_i v_i\over 2}{{\bf B}\over  |{\bf p}-{\bf k}_i|^2}\quad,\nonumber\\
\sqrt{G_-}\dot{{\bf p}}&=& -v_i {({\bf p}-{\bf k}_i)\times {\bf B}\over|{\bf p}-{\bf k}_i|}\quad,\label{holeeq}
\eear
where $\sqrt{G_-}=(1-{\bf b}\cdot {\bf B})$. Note that the term in the first equation that is linear in ${\bf B}$ which leads to the anomaly-induced current has the same sign as in the positive branch.
The contribution to the current from the negative branch then reads as
\bear
{\bf j}_-&=&\int {d^3 p\over (2\pi)^3} f(E) \sqrt{G_-}\dot{\bf x} ={\bf B}{k_i\over 4\pi^2}\int f(E){v_i\over |{\bf p}-{\bf k}_i|^2}|{\bf p}-{\bf k}_i|^2 d|{\bf p}-{\bf k}_i|\nonumber\\&=&
{\bf B}{k_i\over 4\pi^2}\int f(E)v_i d|{\bf p}-{\bf k}_i|={\bf B}{k_i\over 4\pi^2}\int_{E_0}^{E_i} f(E)dE\quad,\label{holeres}
\eear
where in the last equation, we use $E=E_i-v_i|{\bf p}-{\bf k}_i|$ and $E_0$ is the physical cutoff of the bottom of the filled ``Dirac sea'' that one can see in Figure \ref{fig0}. 
At the energy $E_0$, it is intuitively clear that the states from $k=1$ Weyl point meet the states from the other $k=-1$ Weyl point, and become non-chiral ``massive'' Dirac states for which the anomaly-induced transport disappears, see e.g. \cite{Goswami}.
The cutoff is therefore physical. Since $f(E)$ is order 1 around $E_0$, this cutoff is important.
It is natural to assign the common cutoff $E_0$ to the two Weyl points of $k=\pm 1$. 
The distribution $f(E)$ in equilibrium is the same as in (\ref{disteq})
that applies to {\it all} energies, irrespective of branches. 

Expressions (\ref{holeres}) with (\ref{parres}) have the same form and differ only by the distribution function $f(E)$.
To see that this is a correct result, let us check it for the case of relativistic chiral fermions.
The negative branch in that case is the ``Dirac sea'' of negative energy which should be filled in the vacuum state.
Using the identity 
\be
f(-|E|)=1-\bar f(|E|)\quad,
\ee
where $\bar f$ is the anti-particle distribution, and subtracting 1 from the above precisely corresponds to subtracting the vacuum contribution since $f_{\rm vac}(-|E|)=1$, the {\it vacuum-subtracted} negative branch contribution reads as
\be
{\bf B}{k_i\over 4\pi^2}\int_{-\infty}^{0}(- \bar f(|E|))dE=-{\bf B}{k_i\over 4\pi^2}\int_{0}^{\infty} \bar f(E)dE\quad,
\ee
which is the usual negative contribution from anti-particles.

In our condensed matter case, there is nothing to subtract since the ``Dirac sea'' is physical; the ``vacuum'' contribution is very important to keep.
Summing (\ref{holeres}) and (\ref{parres}), the total result simplifies as
\be
{\bf j}={\bf j}_+ +{\bf j}_- = {\bf B}{k_i\over 4\pi^2}\int_{E_0}^{\infty} f(E)dE\quad,
\ee
which is a single integral of $f(E)$ from the bottom of the filled sea to the high energy cutoff.
Note that there is no dependence left on the energy $E_i$ of the Weyl point since $f(E)$ in equilibrium does not depend on it. Using (\ref{disteq}) we can evaluate the expression above in equilibrium as
\be\label{cme1}
{\bf j}_i={\bf B}{k_i\over 4\pi^2}{1\over\beta}{\rm log}\left(1+e^{-\beta(E_0-\varepsilon_F)}\right)\quad,
\ee
with the zero temperature limit
\be\label{cme2}
{\bf j}_i(T=0)={\bf B}{k_i\over 4\pi^2}\left(\varepsilon_F-E_0\right)\quad.
\ee

To summarize this discussion, the chemical potential that enters the formula (\ref{current}) should be measured
{\it from the bottom of the filled sea} $E_0$, not from the Weyl points.
It is clear from the above result that after summing over all Weyl points, the net chiral magnetic current
vanishes in global equilibrium where each Weyl point is filled up to the same Fermi energy, since $\sum_{i} k_i =0$. This is true at any temperature. 
The above formula also indicates that the chiral magnetic conductivity depends non-trivially on the temperature.
This is understandable since we have a physical cutoff ($E_0$) for the bottom of the filled sea.

At this point, we would like to elaborate on the existence of the chiral magnetic effect in more detail. In particular, there are several studies that have found a non vanishing chiral magnetic current in Weyl semimetals. For example in  \cite{cme,zyuzin_burkov,Goswami,cwb} a non zero chiral magnetic current arises in an effective theory where there is an effective background axion field with a non-zero gradient $b^\mu$, identified by the energy-momentum separation of the Weyl points. In our kinetic theory approach,  the chiral magnetic current also exists, but only provided  that left- and right-handed sectors  are filled up to different energies as depicted in Fig. \ref{fig0} (b). In this case, each node contributes to the current by an amount 
\be
{\bf j}_i={\bf B}{k_i\over 4\pi^2}\left(E_i-E_0\right)\quad,
\ee
leading to the net current 
\be
{\bf j}={\bf j}_++{\bf j}_-={{\bf B}\over 4 \pi^2}(E_2-E_1)\quad.
\ee

\subsection{Chiral vortical effect }
We can repeat similar steps for the derivation of the chiral vortical effect, following the suggestion in Ref.\cite{Stephanov:2012ki}
that the rotation can be included as a Coriolis force in the rotating frame;
the equation for $\dot{\bf p}$ becomes 
\be
\dot{\bf p}= \pm 2v_i|{\bf p}-{\bf k}_i| {\vec \omega}\times\dot{\bf x}= 2\left(E-E_i\right){\vec \omega}\times\dot{\bf x}\quad,
\ee
where $\pm$ is for positive and negative branches respectively, and the last equation is true irrespective of branches.
We restrict our attention to the case where $\vec \omega $ and ${\bf k}_i$ are parallel so that Weyl points remain static 
in the rotating frame.
The equation above implies that the magnetic field $\bf B$ can be replaced by \cite{Stephanov:2012ki}
\be
{\bf B}\to 2 \vec \omega \left(E-E_i\right)\quad,
\ee
which leads to the final result
\be\label{cvc1}
j_i=\vec\omega {k_i\over 2\pi^2} \int_{E_0}^\infty dE\,\left(E-E_i\right)f(E)\quad,
\ee
with the zero temperature limit
\be
j_i=\vec\omega {k_i\over 4\pi^2}\left(\varepsilon_F -E_0\right)\left(\varepsilon_F+E_0-2 E_i\right)\quad.
\ee
After summing over all Weyl points, the net current is
\be\label{cvc2}
j_{\rm total}=-\vec\omega {\sum_i k_i E_i\over 2\pi^2}\left(\varepsilon_F-E_0\right)\quad,
\ee
which may {\it not} be zero even in equilibrium.
In this case, the necessary energy when we apply a parallel electric field can be provided by the external rotation -- so the argument that we gave at the beginning of the section does not apply.

\section{Anomaly cooling of Weyl semimetal}

\begin{figure}[t]
	\centering
	\includegraphics[scale=0.7]{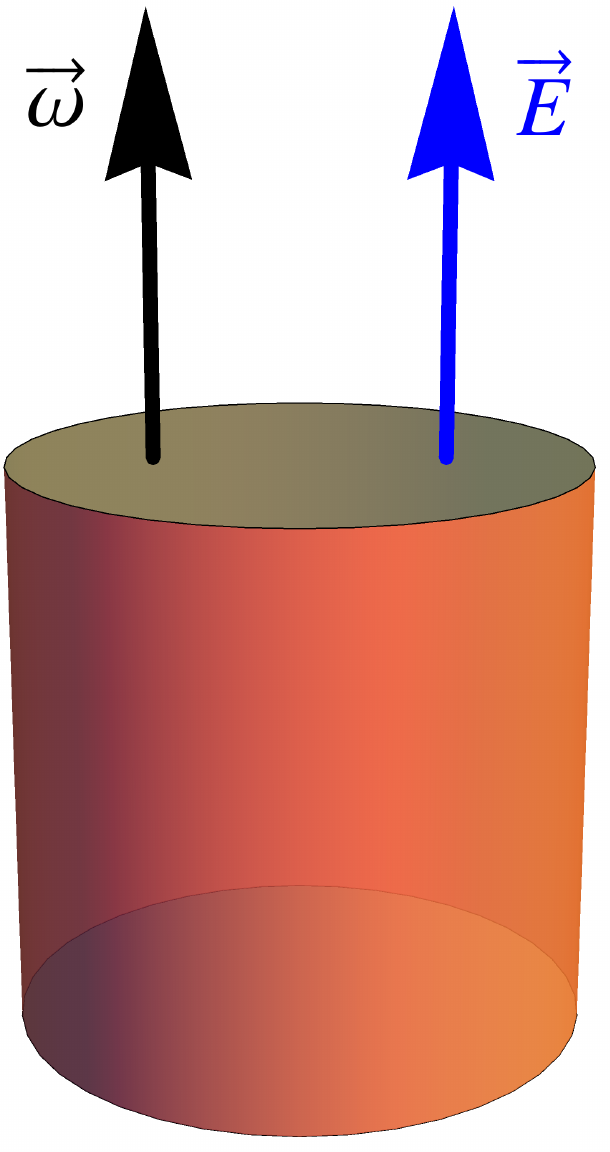}
		\caption{Anomaly cooling of Weyl semimetal by rotation and electric field.\label{fig1} The semimetal could be either cooled or heated depending on the relative orientation of electric field and the angular momentum.}
\end{figure}

Quantum anomaly in Weyl semimetals leads to an interesting phenomenon - it appears that one may use the 
combination of rotation and external electric field to cool this material. To see this, let us use the 
setup explained in Figure \ref{fig1}. 
We rotate the Weyl semimetal in a two dimensional plane, say $(x^1,x^2)$, with angular velocity $\omega$, so that
$\vec\omega =\omega \hat x^3$.
From (\ref{current}), this induces the current along $x^3$ direction as
\bear
J^3={e\omega \over 4\pi^2}\sum_{i} {k_i}\left(\mu_i^2+{\pi^2 \over 3}T^2\right) \quad,
\eear
where we sum over all Weyl points of the system labeled by index $i$.
We then apply an external electric field along $x^3$ direction $\vec E=E \hat x^3$, and
the total power the system absorbs will be given by
\bear
{d {\cal E}\over dt}=\vec J\cdot \vec E = {e\omega E \over 4\pi^2}\sum_{i} {k_i}\left(\mu_i^2+{\pi^2 \over 3}T^2\right) \quad,\label{power}
\eear
which can be either positive or negative depending on the sign of $\omega E$. If it is negative the system will be cooled down.

More intuitively, we can understand the cooling in terms of the entropy current (\ref{entropy}). We will now show that
there exists an  entropy flow directed radially outwards from the system $\vec S=S_r \hat r$, and since no net entropy is generated by the anomaly-induced transport, the entropy of any bounded region around the center should decrease and the system should indeed cool down.
For a finite-size system with radius $R$, the entropy extracted from the cooled central
region will accumulate around the boundary $r=R$, causing a temperature gradient between the center and the boundary
and a compensating heat flow will develop. The system will eventually reach an equilibrium with a stationary temperature gradient
along the radial direction.  

Having a constant $\vec \omega =\omega \hat x^3$ means that the fluid at a position $(x^1,x^2)=r \hat r$ has
a tangential velocity (counter-clockwise) $\vec v=r\omega \hat t$, where $\hat r$ and $\hat t$ are radial and tangential
unit vectors respectively. In the local rest frame of that fluid cell, the fluid experiences a radial magnetic field
\be
\vec B=\vec E\times\vec v=- \omega E r \hat r\quad,
\ee
via Lorentz transformation of field strengths. By (\ref{entropy}) this gives a radial entropy flow
\be
\vec S = e\sum_i k_i\left({1\over 8\pi^2}{\mu_i^2\over T}+{T\over 24}\right)\vec B=
-e\omega E r \sum_i k_i\left({1\over 8\pi^2}{\mu_i^2\over T}+{T\over 24}\right)\hat r\quad,
\ee
with divergence given by
\be
\vec\nabla\cdot \vec S = {1\over r}\partial_r \left(r S_r\right)=-2e\omega E \sum_i k_i\left({1\over 8\pi^2}{\mu_i^2\over T}+{T\over 24}\right)=-{e\omega E\over 4\pi^2 T}\sum_i k_i\left({\mu_i^2}+{\pi^2\over 3}T^2\right)\quad.
\ee
Since no net entropy should be produced, we have ${dS\over dt}+\vec\nabla\cdot \vec S=0$ where $S$ is the entropy density of the fluid; this tells us that the local entropy density changes as
\be\label{entrop}
{dS\over dt}=-\vec\nabla\cdot \vec S={e\omega E\over 4\pi^2 T}\sum_i k_i\left({\mu_i^2}+{\pi^2\over 3}T^2\right)\quad.
\ee
The relation $d{\cal E} = TdS$ precisely reproduces the previous power formula (\ref{power}) from the above. 

Using the fact that $\sum_i k_i=0$ for Weyl semimetals, the $T^2$ term in the cooling rate drops in the final result,
and one needs an asymmetric distribution of $\mu_i$'s to get a non-vanishing effect.
This can be achieved by shifting the energy of Weyl points, as discussed previously. 
Note also that applying the electric field would induce the ordinary current $\sigma \vec E$ where $\sigma$ is the conductivity, which leads to a dissipative heating of the system
\be
{d{\cal E}\over dt}=\sigma E^2\quad,
\ee
which is quadratic in $E$. For the anomaly cooling, which is linear in $E$, to dominate over this dissipative heating, one therefore needs a smaller $E$ and a larger $\omega$.

\section{Charge transport in rotating ``hot'' Weyl semimetal and the Burgers equation}

\begin{figure}[t]
	\centering
	\includegraphics[scale=0.7]{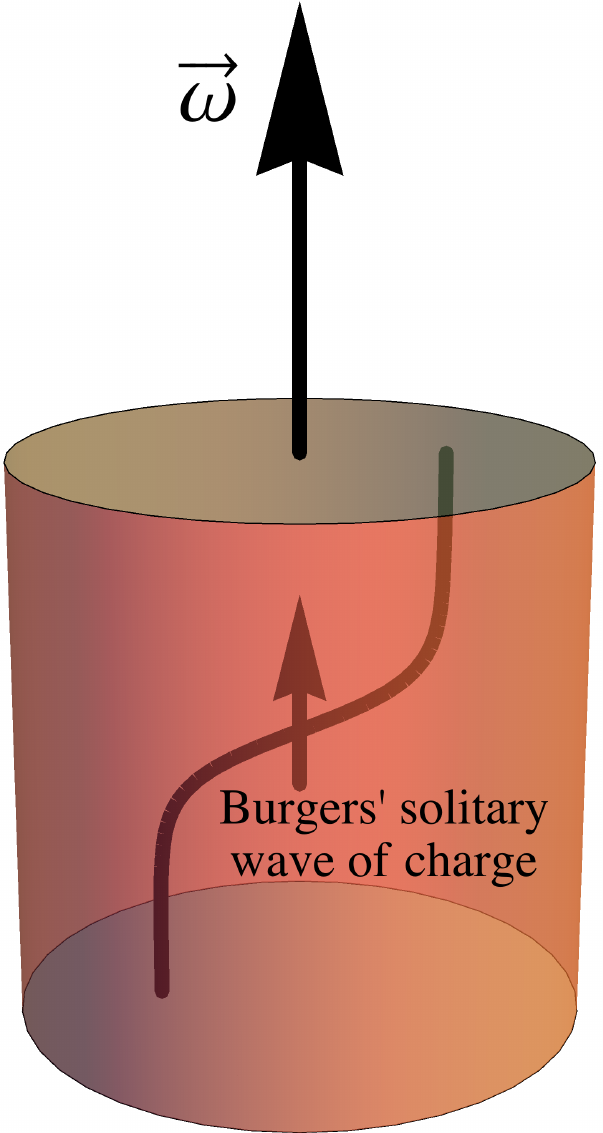}
		\caption{Anomaly induced solitary wave of charge governed by the Burgers' equation.\label{fig2}}
\end{figure}

Let us consider the  long wavelength collective charge transport in a rotating Weyl semimetal as in Figure \ref{fig2}.
Our starting point is again (\ref{current}) with a fixed angular momentum $\vec \omega =\omega \hat x^3$
\be
J^3={k e \over 4\pi^2}\left(\mu^2+{\pi^2 \over 3}T^2\right) \omega\quad.
\ee
Although the total current is the sum over all Weyl points, one can treat contributions from each Weyl point independently, to a good approximation; anomaly-induced collective charge transports from each Weyl point behave independently of other Weyl points.
In more explicit terms, one can introduce $U(1)_i$ charge symmetry for each $i$'th Weyl point separately, and these
$U(1)_i$'s are approximately conserved.
Each $U(1)_i$ has its own triangle anomaly with coefficient $k_i$, and its anomaly-induced current $\vec J_i$ is given by (\ref{current})
with $k\to k_i$ and $\mu\to\mu_i$. 
Note that this is a non-trivial statement because ordinarily the quasi-particles from different Weyl points interact with each other, and this interaction affects their normal (non-anomalous) transport properties.
What protects the independency of anomaly-induced transports between different Weyl points is the triangle anomaly of
$U(1)_i U(1)_j U(1)_k$ which is not zero only if $i=j=k$.

An analogy to a massless quark in QCD may be helpful: there we have two Weyl points, one left-handed ($k=-1$)
and the other right-handed ($k=1$). The left-handed fermions interact strongly with the right-handed fermions in general,
and their normal transport properties are not independent at all. Yet, their anomaly-induced transports are independent:
\be
\vec J_{L,R}=\mp{ e\over 4\pi^2 } \mu_{L,R} \vec B\mp{ 1 \over 4\pi^2}\left(\mu_{L,R}^2+{\pi^2 \over 3}T^2\right) \vec \omega\quad.
\ee
This feature is dictated by the two
separate triangle anomalies of $U(1)_L^3$ and $U(1)_R^3$ without crossing. This independency of collective charge transports of two chiralities is the essence of
chiral magnetic waves proposed in Ref.\cite{Kharzeev:2010gd,Newman:2005hd}. Weak residual interactions between these two chiral collective transports (analogs of sphalerons in the case of QCD) lead to diffusive chiral magnetic waves.

Let us therefore restrict ourselves to a single Weyl point, omitting $i$, with anomaly-induced current at fixed $\omega$ is
\be
J^3={k e \over 4\pi^2}\left(\mu^2+{\pi^2 \over 3}T^2\right)\omega-D\partial_3 \rho +{\cal O}\left(\partial^2\right)\quad,\label{current2}
\ee
where we include the usual diffusion term in the derivative expansion up to first order with diffusion coefficient $D$.
We focus on the regime ${\mu\over T}\ll 1$ of a ``hot'' Weyl semimetal. In this case, $\mu$ is approximately proportional to the charge density $\rho$ by
\be
\mu \approx\chi^{-1} \rho+ {\cal O}\left(\rho^3\right)\quad,\label{amp}
\ee
with susceptibility $\chi$, so that the (\ref{current2}) becomes
\be
J^3={k e \over 4\pi^2}\left(\chi^{-2}\rho^2+{\pi^2 \over 3}T^2\right)\omega-D\partial_3 \rho +{\cal O}\left(\partial^2,\rho^3\right)\quad.
\ee
Using this in the charge conservation equation $\partial_t \rho+\partial_3 J^3=0$ leads to
\be\label{burgers}
\partial_t\rho+C\rho\partial_x\rho-D\partial_x^2\rho=0\quad,\quad x\equiv x^3\quad,
\ee
with $C={k e\omega\over2\pi^2\chi^2}$. This is the Burgers' equation, a prototypical integrable partial differential equation in 1+1 dimensions. It is completely soluble given an initial data, and many analytic solutions are known.
Shockwave-type solutions are not acceptable in our problem because of our small amplitude approximation (\ref{amp}),
but solitary traveling waves with finite amplitudes are relevant. One can tune the constant $C$ by varying $\omega$.

The total electromagnetic current is given by the sum over each Weyl point contributions $\sum_i \vec J_i$.
In practice one may perturb the system by injecting a net electromagnetic charge, and it may be hard to excite charge fluctuations of each $U(1)_i$ individually; one generally excites a superposition of all $i$'th charges. Since each $i$'th fluctuations are
propagating independently, one would observe a splitting of charge transport. This is very interesting as it would allow to study
the properties of each Weyl point separately; charge transport is a prism for probing different Weyl points in a Weyl semimetal.

\section{Quantized vortices in a Weyl superfluid}

Let us now move to a slightly different topic and consider a superfluid system which couples to Weyl fermions as in \cite{Kirilin:2012mw,Zakharov:2012vv}. This type of configuration was studied in \cite{Kogut:1999iv,Son:2000by} as a phenomenological description of quark matter at finite isospin chemical potential. Furthermore, it was recently observed that Weyl fermion excitations can be realized in ultra-cold fermionic gases in the presence of Zeeman field and Rashba spin-orbit coupling \cite{Gong, Seo:2013oja}.    
  
  We start with the following Lagrangian:
\be
\mc L=\bar\psi i \gamma^\mu(\mc D_\mu+i \partial_\mu\phi)\psi+\frac{1}{2}|\mc D_\mu \chi|^2
\ee
Here $\chi=|\chi|e^{i\phi}$ is the bosonic field which condenses and forms the superfluid and $\mc D_\mu=\partial_\mu-ieA_\mu$. The phase $\phi$ is related with the superfluid velocity as follows:
\be
u_\mu=\frac{1}{m}\partial_\mu\phi 
\label{velocity}.
\ee
Here $m$ is the scale that fixes the magnitude of $u_\mu$. The superfluid has vortex line configurations which are of the form
\be
\chi=f(x_\perp)\frac{x_1+i x_2}{r}
\ee
where $\vec x_\perp=(x_1,x_2)$ denotes the transverse plane where the fluid rotates and $r=\sqrt{x_1^2+x_2^2}$. The vortex is centered at the origin $x_1=x_2=0$ and is elongated along a line in $x_3$ direction. The function $f(r)$ is constant at large values of r, $f(\infty)\equiv\sqrt{\rho(T)}/m$ and vanishes at the center $r=0$.  The constant value is associated with the superfluid density $\rho(T)$ which decreases with temperature and vanishes above the critical temperature $T_c$. The distance $r_0$, in which $f(r)$ changes from $0$ to its constant value is the radius of the core of vortex. We will assume that $r_0$ is smaller than any macroscopic scale in our problem. 

The identification (\ref{velocity}) leads to the Onsager-Feynman quantization
\be
\int(\vec\nabla_\perp \phi).d\vec l=2 \pi  n\quad,\quad n=0,\pm1,\pm2,...
\ee
The integer $n$ is the winding number of the vortex and is thus a topological invariant. From now on we will focus on a single vortex with unit vorticity  $n=1$ since a vortex with a multiple winding number is unstable against decaying into multiple vortices with unit winding.

Assuming that the distance between the two vortices is much larger than the core radius ($r_{12}>>r_0$) and neglecting the contribution from the cores, the interaction energy between the two vortices with the same vorticity is given by
\be
E_K=2 L\rho \int d^2x_\perp\, \vec u_1 . \vec u_2\approx L\frac{2\pi\rho}{m^2}\ln(r_{12}/r_0)
\ee
 where $L$ is the size of the sample along the vortex line.  
 The chemical potential for fermions can be realized as a shift in the superfluid phase
 \be
 \phi\rightarrow\phi+\mu\,t
 \ee
 In \cite{Kirilin:2012mw} it was shown that the existence of a chemical potential induces a charge current through the triangle anomaly,
 \be
 J^3=\frac{\mu}{2\pi}.
 \ee
 Note that we are considering Weyl fermions, and therefore there is no distinction between the axial current and charge current since there is only a single anomalous U(1) symmetry in our problem.
 This current is localized inside the vortex core which can be treated point-like in $x_1x_2$ plane. The existence of  the anomalous current will modify the interaction between the vortices since the two currents repel each other. The energy of this current-current interaction has the same form as the vortex-vortex interaction
 \be
 E_{cc}=L \frac{J^3_1 J^3_2 }{2 \pi} \ln (\frac{r_{12}}{r_0})=\frac{\mu^2}{8\pi^3}\ln (r_{12}/r_0)
 \ee
As a result, the total interaction energy between the two vortices is modified as
\be
\frac{E}{L}=\left(\frac{2\pi\rho}{m^2}+\frac{\mu^2}{8\pi^3}\right)\ln \left(\frac{r_{12}}{r_0}\right)
\label{vort_int}
\ee

This anomalous contribution to the vortex dynamics has interesting consequences. For example, consider a very thin sample, such as a thin film which can be treated effectively as two dimensional. In this case, above a certain temperature $T_{BKT}$ creation of vortices becomes energetically favorable. This is the famous Berezinskii-Kosterlitz-Thouless  (BKT) transition \cite{BKT}. We now show that the existence of the anomalous current modifies the BKT transition. Following the standard argument, let us calculate the free energy of a single vortex configuration
  \be
 F=E-T S
 \ee
where $S$ is the entropy. If we assume that the vortices are distributed in the two dimensional plane of the thin film of area $R^2$ much larger than the vortex size, we can have $R^2/r_0^2$ possible configurations and the entropy is simply $S=\ln(R^2/r_0^2)$. The energy $E$ has two terms, the kinetic term due to the velocity field and the magnetic energy induced by the anomalous current:
\be
E=\frac{1}{2}L\int d^2x_\perp (\vec u^2+\vec B^2)=\pi L\int_{r_0}^R r dr \left(\frac{\rho}{r^2m^2}+\frac{\mu^2}{4 \pi^2 r^2}\right)\approx \left(\frac{L\pi\rho}{m^2}+\frac{L\mu^2}{4\pi^3}\right)\ln (R/r_0)
\ee
We again neglected the contribution of the core which is valid in the thermodynamic limit $R>>r_0$. The free energy in this limit is
\be
F=\left(\frac{L\pi\rho}{m^2}+\frac{L\mu^2}{4\pi^3}-2T\right)\ln (R/r_0)
\ee
The phase transition occurs when the free energy changes sign and when it is negative it is preferable to create vortices. The phase transition temperature is
\be\label{bkt}
T_{BKT}=\frac{L\pi}{2m^2}\,\rho(T_{BKT})+\frac{L\mu^2}{8\pi^3}
\ee
The first term is the well known expression for the temperature of the BKT phase transition, whereas the second term is the modification due to the anomaly. It is possible to consider the full dynamics by taking into account the screening of the vortices and solving the gap equation as in \cite{BKT}. In the thermodynamic limit the vortex interaction (\ref{vort_int}) will be made stronger by the additional anomalous term with the same spatial dependence -- therefore the effect of anomaly can be included in a straightforward way.

\section{Summary}

Let us briefly summarize our results:

i) The existence of the Chiral Magnetic Effect (CME) in Weyl semimetals depends on how the left- and right-handed sectors are filled. We found that if each sector is filled up to the same Fermi energy, the CME vanishes. However, if each sector is filled up to a different energy, and the resulting chiral chemical potential can evolve in time, the current exists -- see Figure \ref{fig0}. The latter configuration can be realized, for example, when the initial chiral chemical potential is induced dynamically, e.g. by parallel electric and magnetic fields. Furthermore, the current also exists in the presence of \textit{time dependent} magnetic fields, in agreement with \cite{cwb}. The chiral magnetic current is given by (\ref{cme1}) and by (\ref{cme2}) for the case of zero temperature.

ii) Contrary to the chiral magnetic effect, the chiral vortical effect in Weyl semimetals can exist even in equilibrium, with time-independent chiral chemical potentials. The chiral vortical current is given by (\ref{cvc1}) and by (\ref{cvc2}) for the case of zero temperature.

iii) The chiral vortical effect in a rotating Weyl semimetal leads to a very interesting phenomenon -- the ``anomaly cooling", when the temperature of the material with a non-zero chiral chemical potential can be reduced as a result of rotation. The local entropy density changes according to (\ref{entrop}). 

iv) The anomaly-induced transport of charge in rotating ``hot" (with chemical potential much smaller than temperature) Weyl semimetals is described by the integrable Burgers' equation (\ref{burgers}) that 
admits solitary wave solutions.

v) The anomaly induces a new term in the interaction energy of quantized vortices in a superfluid coupled to Weyl fermions. This shifts the energy of the BKT phase transition according to  (\ref{bkt}).

Quantum anomalies are among the most subtle and beautiful effects in relativistic field theory.
Weyl semimetals open an intriguing possibility to study the effects of quantum anomalies experimentally, in a controlled setting. Such studies are of fundamental interest, and may lead to practical applications as well.

\vskip 2cm 
\centerline{\large \bf Acknowledgements} \vskip 0.5cm

This work was
supported by the U.S. Department of Energy under Contracts No. DE-FG-88ER40388 and
DE-AC02-98CH10886 (GB and DK). We thank L. Levitov, M. Stephanov, Y. Yin and I. Zahed for useful discussions.

 \vfil

\end{document}